\documentclass{emulateapj}
\usepackage{apjfonts}

\shorttitle{Off-axis X-ray transients from short GRBs}
\shortauthors{Lazzati et al.}

\begin{document}

\title{Off-axis prompt X-ray transients from the cocoon of short
  gamma-ray bursts}

\author{Davide Lazzati$^1$, Diego L\'opez-C\'amara$^2$, Matteo
  Cantiello$^{3,4}$, Brian J. Morsony$^5$, Rosalba Perna$^6$, Jared
  C. Workman$^{7}$}

\affil{$^1$ Department of Physics, Oregon State University, 301
  Weniger Hall, Corvallis, OR 97331, USA} 

\affil{$^2$ CONACYT -- Instituto de Astronom\'ia, Universidad Nacional
  Aut\'onoma de M\'exico, A.P. 70-264, 04510 M\'exico D.F., M\'exico}

\affil{$^3$ Center for Computational Astrophysics, Flatiron Institute,
  162 5th Ave, New York, NY 10010}

\affil{$^4$ Dept. of Astrophysical Sciences, Peyton Hall, Princeton
  University, Princeton, NJ 08544, U.S.A.}

\affil{$^5$ Department of Astronomy, University of Maryland, 1113
  Physical Sciences Complex, College Park, MD 20742-2421, USA}

\affil{$^6$ Department of Physics and Astronomy, Stony Brook
  University, Stony Brook, NY 11794-3800, USA}

\affil{$^{7}$ Department of Physical and Environmental Sciences,
  Colorado Mesa University, Grand Junction, CO 81501, USA}

\begin{abstract}
  We present the results of numerical simulations of the prompt
  emission of short-duration gamma-ray bursts. We consider emission
  from the relativistic jet, the mildly relativistic cocoon, and the
  non-relativistic shocked ambient material. We find that the cocoon
  material is confined between off-axis angles
  $15\lesssim\theta\lesssim45^\circ$ and gives origin to X-ray
  transients with a duration of a few to $\sim10$ seconds, delayed by
  a few seconds from the time of the merger. We also discuss the
  distance at which such transients can be detected, finding that it
  depends sensitively on the assumptions that are made about the
  radiation spectrum. Purely thermal cocoon transients are detectable
  only out to a few Mpc, Comptonized transients can instead be
  detected by the FERMI GBM out to several tens of Mpc.
\end{abstract}


\section{Introduction}

The discovery of Gravitational Waves (GWs) from mergers of binary
black holes (BHs) has openend a new window of study in the Universe
\citep{Abbottetal2016a,Abbottetal2016b}. Each event has been
accompanied by a massive search for electromagnetic (EM) counterparts,
despite the lack of a general consensus for a production mechanism of
such counterparts to BH-BH mergers
\citep{Connaughtonetal2016,Loeb2016,Pernaetal2016,Zhang2016,DeMinketal2017}.
The identification of an EM counterpart would possibly lead to the
identification of the host galaxy and to a redshift
measurement. Moreover, EM signals carry complementary information that
would allow for better constraints on the source properties.

Unlike BH-BH mergers, the coalescence of two neutron stars (NSs), or a
NS and a BH, is expected to be accompanied by EM radiation: these
events are believed to be the progenitors of short Gamma-Ray Bursts
(SGRBs, \citealt{Eichleretal1989,Nakar2007,Berger2014}). However,
despite decades of indirect evidence pointing to this association,
only a detection of simultaneous GWs and gamma-ray radiation would
constitute a smoking gun to confirm the association, and finally solve
the long-standing mystery of the origin of SGRBs. Additionally, the EM
counterpart would allow for a better localization of the merger sites
and hence help constrain the evolutionary scenario which led to the
binary formation. Combining the parameters inferred from the GW
radiation (i.e. masses) with the energetics inferred from the SGRB EM
emission, further information can be gained on the mass of the ejecta
\citep{Giacomazzoetal2013}.

The question of the simultaneous observability of GWs and EM radiation
is hence of paramount importance.  However, while GWs are only
moderately anisotropic, the $\gamma$-ray emission is likely produced
within a collimated, relativistic outflow, which reduces the
probability of seeing it in association with a GW
event\footnote{Numerical simulations of binary NS mergers also
  indicate the presence of collimation in the magnetic field
  \citep{Rezzollaetal2011,Kawamuraetal2016,Ruizetal2016}.}.  The longer
wavelength emission, if produced by the same relativistic jet as in
the standard afterglow scenario, is also expected to be collimated, at
least until the jet has slowed down to trans-relativistic speeds,
weeks to months after the event \citep{Rossietal2002}. Observations of
SGRBs so far have estimated the average typical opening angle of the
jet to be $\sim16^\circ$ or less
\citep{Fongetal2015,Ghirlandaetal2016}, making the probability of
seeing a jet on axis (and hence a ``standard'' SGRB) less than $10\%$.

Given the above, additional sources of EM emission from NS-NS mergers
become especially relevant. In the optical/near infra-red band, an
important contribution can be provided by the kilo/macronova, a
transient phenomenon triggered by the radioactive decay of r-process
nuclei in the neutron-rich material ejected during the NS-NS merger
\citep{Lietal1998,Metzgeretal2010,Metzgeretal2012,Kasenetal2013,Kasenetal2015,Kawaguchietal2016}.
At all other wavelengths, the best prospects for detection come from
``side'' emission (SE) from the jet\footnote{Unless the merger remnant
  is a long-lived NS, in which case an additional spindown-powered
  transient is expected
  \citep{Yuetal2013,Metzgeretal2014,Siegeletal2016a,Siegeletal2016b}}. Even
though it is significantly weaker than the on-axis emission, SE is
potentially very important when in association with a GW event, since
the distance to which NS-NS events can be detected with advanced LIGO
is only about 200~Mpc even after reaching design sensitivity
($65-115$~Mpc for advanced Virgo; currently the limit being
$80-120$~Mpc for advanced LIGO and $20-60$~Mpc for advanced Virgo,
\citealt{Abbottetal2016c})\footnote{All distances
  orientation-averaged.}. The fainter SE therefore significantly
increases the chances of detectability for off-axis events.

A study of side emission from the jet becomes especially timely during
the GW era. Observer lines of sights at large angles are much more
likely in association with a GW event from a compact binary merger
than are on-axis observations of SGRBs. The probability density
function of LIGO/Virgo detections with respect to the off-axis angle
peaks at about $30^\circ$ \citep{Schutz2011}. The off-axis emission
from long GRB jets has been widely studied and has taken various names
in the literature, such as structured jet (e.g.,
\citealt{Rossietal2002}), off-axis emission (e.g.,
\citealt{Granotetal2005,Salafiaetal2016}), sheath
\citep{Kathirgamarajuetal2017}, cocoon/wide-angle (e.g.,
\citealt{Lazzatietal2017}). Some theories explain the phenomenon of
X-ray flashes (an X-ray dominated sub-class of long GRBs) as the
off-axis emission from collapsars
\citep{Gotthelfetal2006,Yamazakietal2002,Fynboetal2004,Lambetal2005,Sakamotoetal2005,Guidorzietal2009};
alternative thoeries, though, exist \citep{Ciolfi2016}.

In \cite{Lazzatietal2017} (hereafter L17) we presented calculations of
the wide angle emission of SGRBs from compact binary merger
progenitors. This emission arises from a hot cocoon, which forms if
the jet initially propagates through a baryon-contaminated region
surrounding the merger site
\citep{RamirezRuizetal2002,Nakaretal2017}. The semi-analytical
calculations in L17 were made under a number of approximations, such
as the assumption of an isotropic cocoon. However, simulations by
\cite{Gottliebetal2017} have showed that this is not the case.
Additionally, a specific, ad-hoc value of the Lorentz factor,
$\Gamma=10$, had been adopted.

Here we perform a full simulation of the formation of the cocoon and
its interaction with the jet and unshocked ambient medium, using a
nested 3D/2D approach.  Simulations of the dynamical effect of the
merger ejecta on a relativistic jet were pioneered by
\cite{Nagakuraetal2014} and \cite{Murguiaetal2014,Murguiaetal2017}. We
predict the distribution of the isotropic equivalent energy as a
function of the off-axis angle. We then calculate the profile of the
peak photon energy of the photospheric cocoon emission, as well as the
profile of the pulse duration and time delay of the pulse from the jet
launching time as a function of the off-axis angle. Our numerical
methods are described in Sec.~2, and the results are described in
Sec.~3. We summarize and discuss our results in Sec.~4.

\section{Numerical Methods}

The simulation of a SGRB jet was performed with the adaptive mesh
refinement relativistic hydrodynamic (AMR-RHD) code FLASH
\citep{Fryxelletal2000} as modified in \cite{Morsonyetal2007}. The
SGRB jet was simulated as an inflow boundary condition with similar
properties as the jet in L17: $L_j=10^{50}$~erg/s, $\theta_j=16^\circ$,
$t_{\rm{eng}}=1$~s. However, it was injected with $\Gamma_\infty=300$
and already mildly relativistic ($\Gamma_0=5$) at the inner boundary
located at $r_0=10^8$~cm from the merger site. The local merger ejecta
are approximated as an exponentially cutoff wind, with density
profile:
\begin{equation}
n(r)=n_0 \left(\frac{r}{r_0}\right)^{-2}\,e^{-\frac{r}{r_0}}
\end{equation}
where, again, $r_0=10^8$~cm, and $n_0=10^6$~cm$^{-3}$, for a total
ejecta mass of 0.006 solar masses, 0.002 of which are inside the
simulation domain (the remainder being located at $r<r_0$). Such
masses and densities are comparable to the results of numerical
simulations of binary NS and NS-BH mergers
\citep{Kiuchietal2015,Endrizzietal2016,Radiceetal2016,Ciolfietal2017}.

When simulating the emission from short-duration gamma-ray bursts, it
is challenging to balance the need for a large domain (the fireball
becomes transparent at the photospheric radius
$r_{\rm{ph}}\sim10^{13}$~cm), high spatial resolution (the thickness
of the fireball remains of the order of one light second throughout
the expansion), and full dimensionality (3D) in order to avoid the
occurrence of the plug (or butterfly) instability
\citep{Gottliebetal2017}.  Such instability is particularly harmful in
short GRB simulations since it can penetrate the fireball all the way
to the back of the shell, resulting in complete loss of the
relativistic motion.

\begin{figure}
\includegraphics[width=\columnwidth]{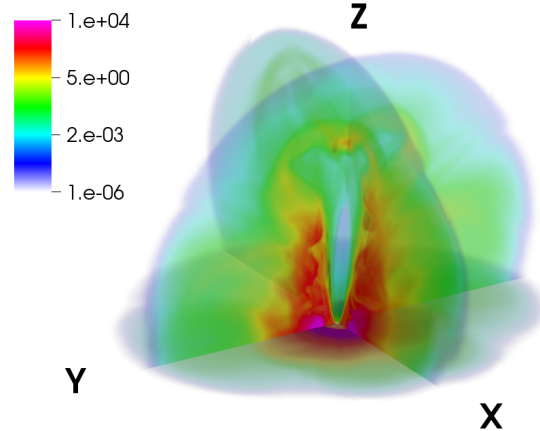}
\caption{{Three dimensional view of the comoving density (g/cm$^3$)
    form our initial 3D simulation of a SGRB jet propagating through
    0.002 solar masses of non-relativistic ejecta. The simulation is
    shown at $t_{\rm{lab}}=0.25$s. The jet vertical size is
    $\sim1.5\times10^{10}$~cm at this time.}
\label{fig:3D}}
\end{figure}

\begin{figure*}
\includegraphics[width=\textwidth]{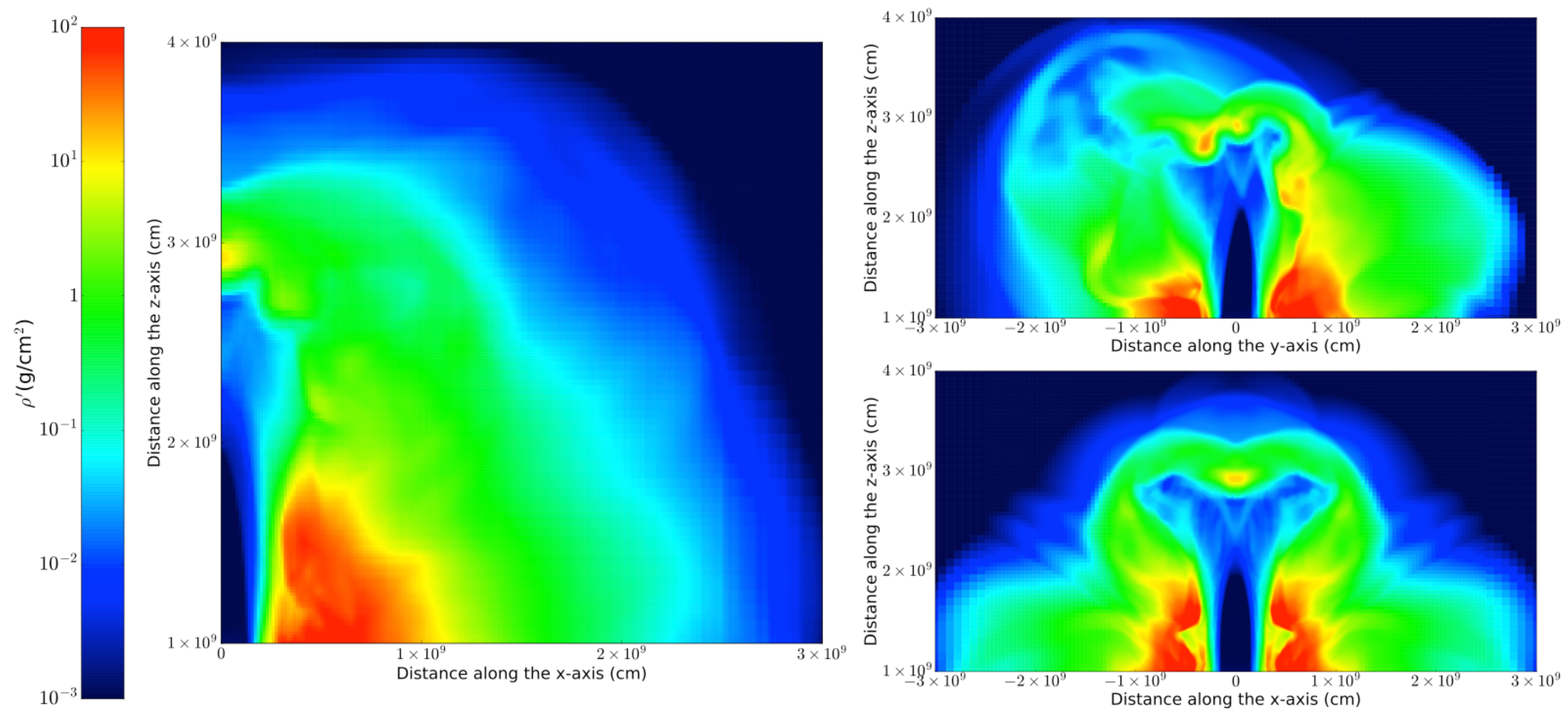}
\caption{{Pseudocolor maps of the logarithmic comoving density of the
    3D simulation at the time of the 3D to 2D projection. The leftmost
    bar shows the color scale and density cuts applied to all the
    images. The left panel shows the azimuthal projection of the 3D
    density in 2D cylindrical coordinates. The right panels show the
    density along the YZ (top) and XZ (bottom) planes.}
\label{fig:3Dpanels}}
\end{figure*}

\begin{figure*}
\includegraphics[width=0.5\textwidth]{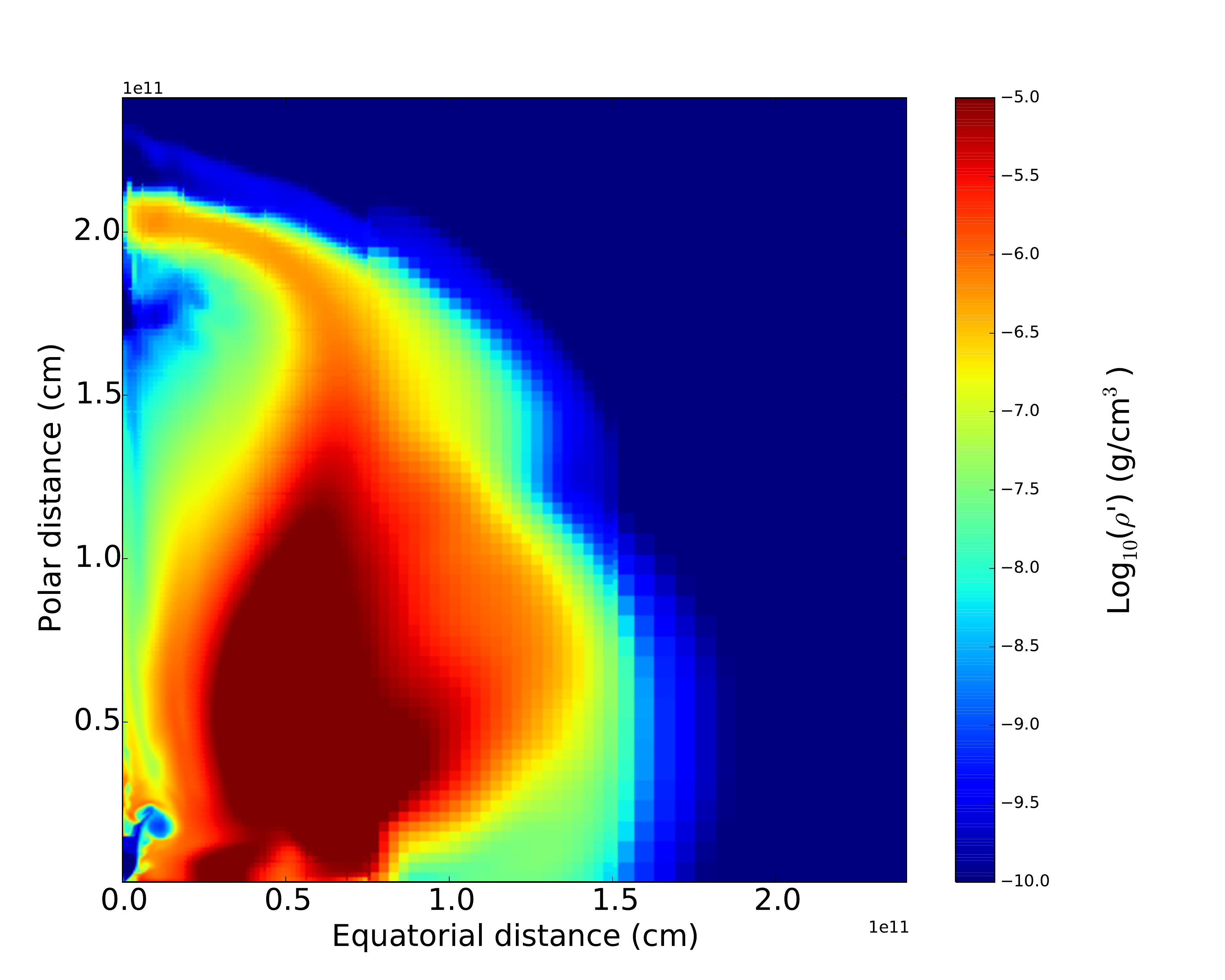}
\includegraphics[width=0.5\textwidth]{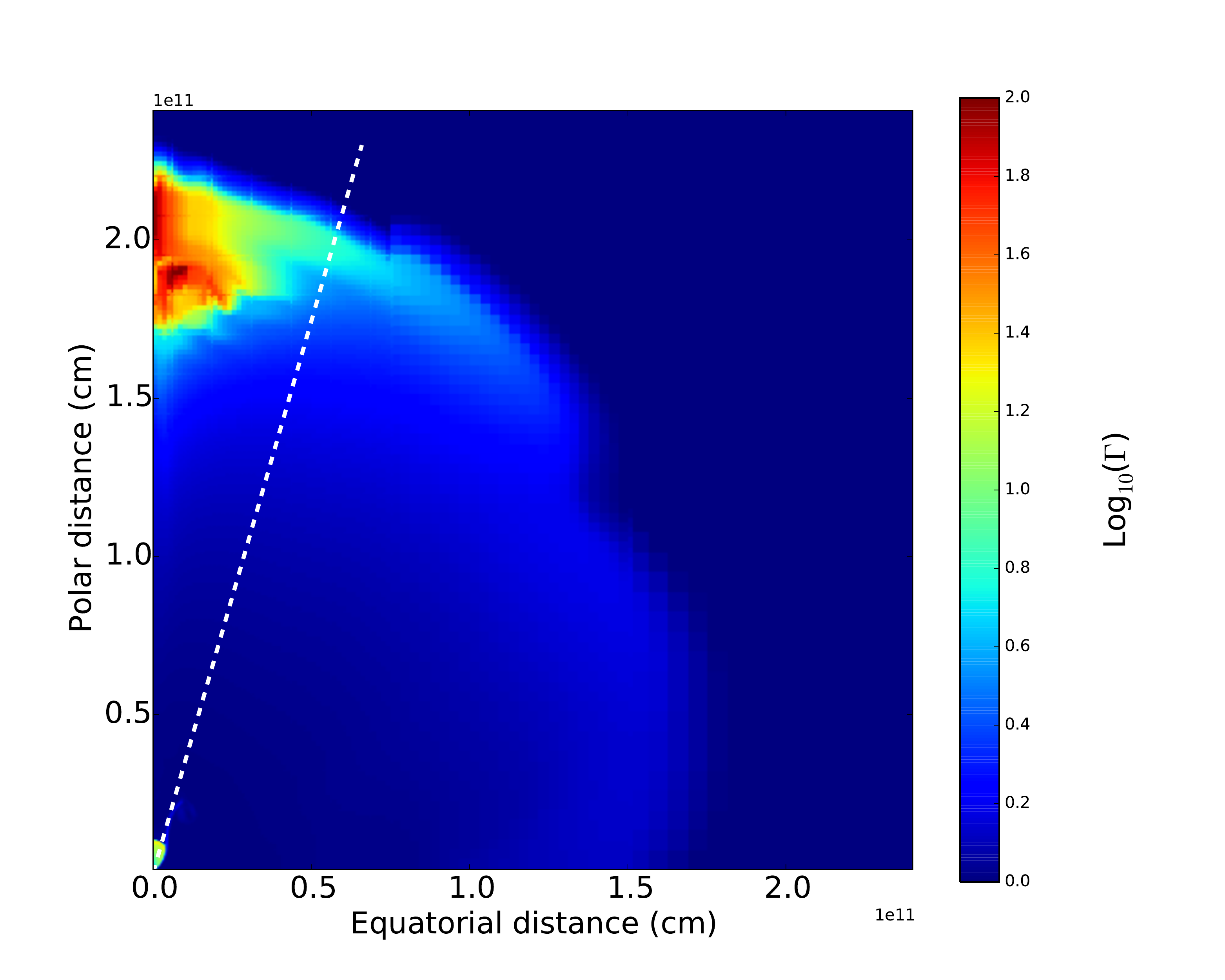}
\caption{{Pseudocolor map of the logarithm of comoving density (left
    panel) and Lorentz factor (right panel) for our 2D FLASH RHD
    simulation of a short GRB jet. The simulation is shown at
    laboratory time $t_{\rm{lab}}=6.7$ seconds after the jet is
    launched. A white dashed line in the right panel shows the angle
    ($\theta_0=16^\circ$) within which the jet is injected at the
    inner boundary ($r=10^8$~cm).}
\label{fig:2D}}
\end{figure*}

With current technology, it is impossible to perform a simulation in
3D with the required domain and resolution. We therefore proceeded as
follow. We simulate the first 0.25 seconds of the jet evolution in 3D
in a domain $-2\times10^{10} \le x,y \le 2\times10^{10}$~cm,
$10^8 \le z \le 4.01\times10^{10}$~cm.  Our finest grid has a
resolution of $2.4\times10^6$~cm. The density rendering of the 3D
simulation at $t=0.25$~s is shown in Figure~\ref{fig:3D} (where the
jet has reached a size $\sim1.5x10^{10}$~cm). We then project the last
simulation box in 2D by performing an azimuthal average around the
z-axis and use such projection as the initial condition for the
subsequent evolution in 2D cylindrical coordinates (x,z). More detail
about this procedure will be given in an upcoming publication
(Lopez-Camara et al. in preparation).  Figure~\ref{fig:3Dpanels} shows
some detail of the projection procedure. With identical color scale
(shown in the left of the figure), the three panels show the
azimuthally averaged density map (left panel) and the density map in
the YZ and XZ planes of the full 3D simulation (top and bottom right
panels, respectively). The azimuthally averaged properties of the
outflow (such as the density shown in the left panel of
Figure~\ref{fig:3Dpanels} are used as initial conditions for the
subsequent 2D evolution. The 2D simulation box extends from $10^9$ to
$10^{13}$~cm in the polar direction and from 0 to $10^{13}$~cm in the
equatorial plane. Joining the two simulations, we cover 5 orders of
magnitude in scale, from $10^8$ to $10^{13}$~cm, much more than
previously accomplished
\citep{Gottliebetal2017,Kathirgamarajuetal2017}. In terms of spatial
resolution, the 2D simulation has a square of side
$2.44\times10^{7}$~cm at the inner boundary, yielding the same
relative resolution as the 3D simulation. At larger distances, the
highest resolution element becomes a box of side
$\sim1.56\times10^9$~cm, sufficient to fully resolve the SGRB shell of
thickness $\sim3\times10^{10}$~cm. Pseudocolor images of the
logarithmic density and Lorentz factor of the 2D simulation at
$t_{\rm{lab}}=6.7$~s are shown in Figure~\ref{fig:2D}.

To compute the light curve we followed the method of Lazzati et
al. (2009; 2011; 2013, in particular Eq. 2 in Lazzati et al. 2013),
considering only material moving with a speed of at least $0.4$~c,
corresponding to $\Gamma>1.1$. Despite the extended size of the
simulation domain, we were not able to simulate the fireball all the
way to its photosphere, and consequently we had to extrapolate some of
the outflow properties to the desired radii. In particular, if we
measured a comoving temperature $\bar{T}^\prime$ at a radius $\bar{r}$
at which the opacity to Thomson scattering is $\bar\tau\gg1$, we
compute the photospheric radius as
$r_{\rm{ph}}=\bar{r} \sqrt{\bar\tau}$, and the photospheric
temperature as
$T_{\rm{ph}}^\prime=\bar{T}^\prime\bar\tau^{-1/3}$. These scalings are
correct as long as the outflow is in the ballistic regime and the
ejecta coast at constant Lorentz factor in a self-similar fashion. To
check that the extrapolation does not affect the accuracy of the
results, we performed the extrapolation from the simulation results at
four different times: $t_{\rm{lab}}=3.3, 6.7, 13.3$, and $26.6$~s. The
four values of the photospheric radius, energy, and comoving
temperature that we obtained were consistent with each other, yielding
an a-posteriori proof that the ballistic assumption is correct for our
simulations (with some exceptions, see below). In
Figures~\ref{fig:lumi},~\ref{fig:epk}, and~\ref{fig:T90}, we plot the
average of the four values with symbols and show with an error bar the
dispersion of the four measurements. Finally, we adopt some correction
for the computed peak frequency and energy. As shown in Lazzati (2016,
see also Parsotan \& Lazzati 2017), the fluid temperature measured at
the photosphere underestimate by a factor $\sim3$ the temperature of
the radiation, while the energy is overestimated by a factor that
ranges between a few and 10. We therefore correct our photon peak
energies upwards by a factor 3 and our luminosities downward by a
factor 4. More accurate results that make use of the Monte Carlo
radiation transfer code MCRaT will be presented in a future
publication.

\section{Results}

\begin{figure}
\includegraphics[width=\columnwidth]{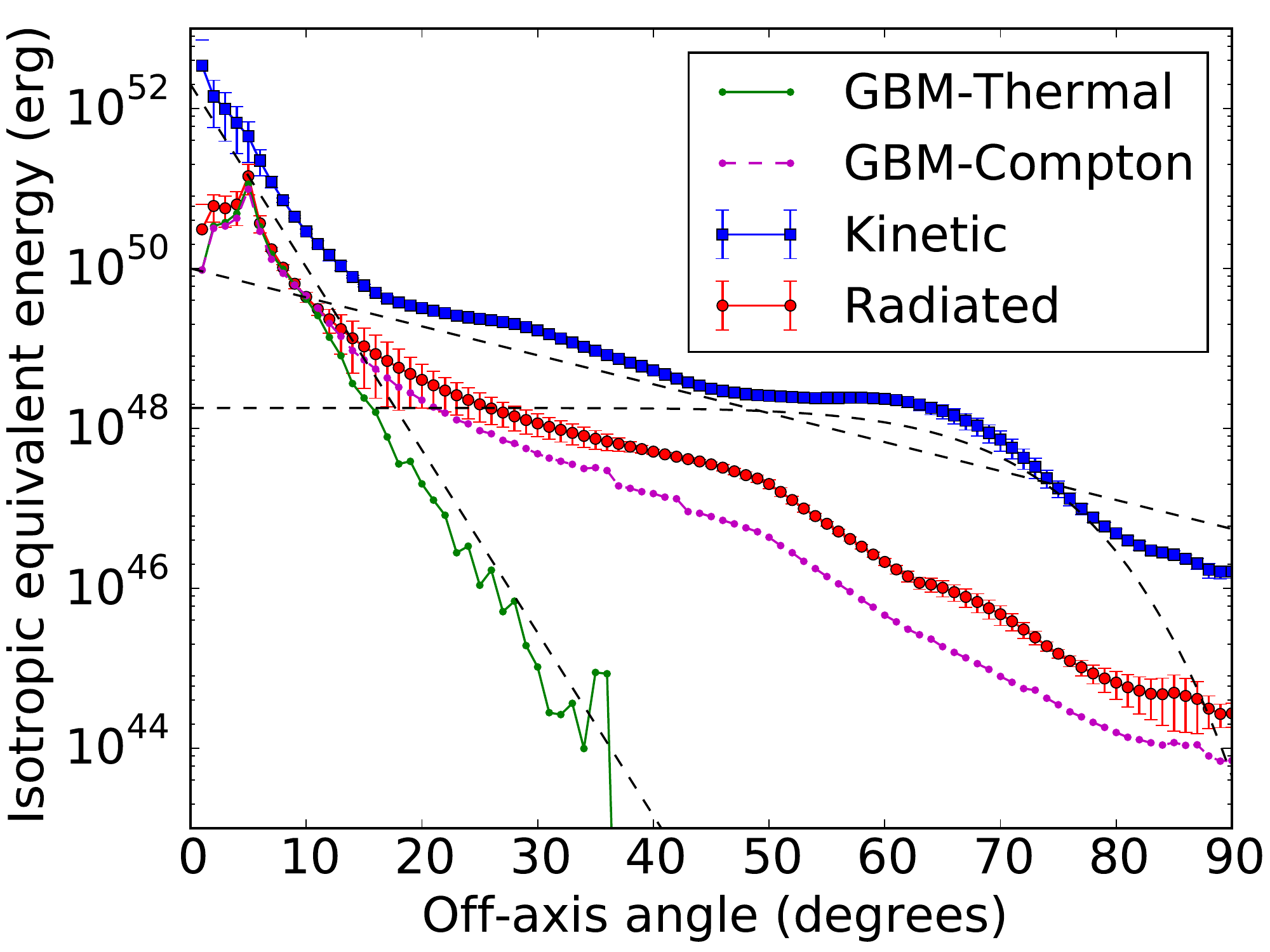}
\caption{{Profile of the isotropic equivalent energy as a function of
    the off-axis angle from our simulation. The energy has been
    computed at several stages of the evolution (see text). The
    symbols show the average result, while the error bars show the
    full range of variation at each angle. Blue square symbols show
    the kinetic energy, while red dots show the bolometric
    energy. Lines with dots shows the energy that would be detected
    within the FERMI GBM sensitivity band for either a pure thermal
    spectrum (green solid line) or a Comptonized spectrum (magenta
    dashed line). Black dashed lines are overlaid on the kinetic
    energy profile to show the three components described in the text:
    the stratified jet (exponential), the cocoon (exponential) and the
    shocked ambient medium (constant with sharp cutoff).}
\label{fig:lumi}}
\end{figure}

Figure~\ref{fig:lumi} shows the polar profile of the kinetic energy of
the ejecta (blue square symbols) and of the isotropic equivalent
energy in radiation (red dots, bolometric). As noted above, the
symbols show the average value obtained for the four different
starting points of the extrapolations and the error bars indicate
their dispersion (one standard deviation). We notice three structures
in the polar profile that we interpret as follows. The brightest part
is the core of the outflow, for $\theta<15^\circ$. This is the
original injected jet, whose energy profile has been modified by the
interaction with the environment from a top-hat jet to an
exponentially stratified structure with e-folding angle
$\theta_{j,{\rm{obs}}}=2^\circ$. This exponential jet is surrounded by
a hot bubble that dominates at angles between 15 and 45 degrees. We
interpret it as the jet cocoon mixed with the ambient material cocoon
\citep{Nakaretal2017}. It is also characterized by an exponential
profile with e-folding angle
$\theta_{\rm{cocoon,obs}}=12^\circ$. Finally, there is a fairly
isotropic component with a sharp cutoff at $\theta=65^\circ$, that we
interpret as the ambient medium shocked by the cocoon pressure while
the jet is still trapped inside the non-relativistic ejecta from the
merger. Whether the cutoff at $\sim 65^\circ$ is physical is unclear
since the reflecting boundary condition at $\theta=90^\circ$ might
affect the hydrodynamical evolution at large angles. These three
components are shown in Figure~\ref{fig:lumi} with dashed lines
overlaid on the kinetic energy curve. In terms of energetics, of the
$10^{50}$ erg injected in the boundary condition, $5.5\times10^{49}$~erg
remain in the jet ($\theta\le15^\circ$), $3.8\times10^{48}$~erg are
found in the cocoon ($15<\theta\le45^\circ$), and $7\times10^{47}$ are
in the shocked ambient medium. The remainder of the energy is given to
slow ejecta ($\Gamma<1.1$). Analogous components could explain the
photospheric radiated energy profile, but the cocoon radiation
requires a smaller e-folding angle to reproduce the steeper decline of
the radiative energy with angle.

\begin{figure}
\includegraphics[width=\columnwidth]{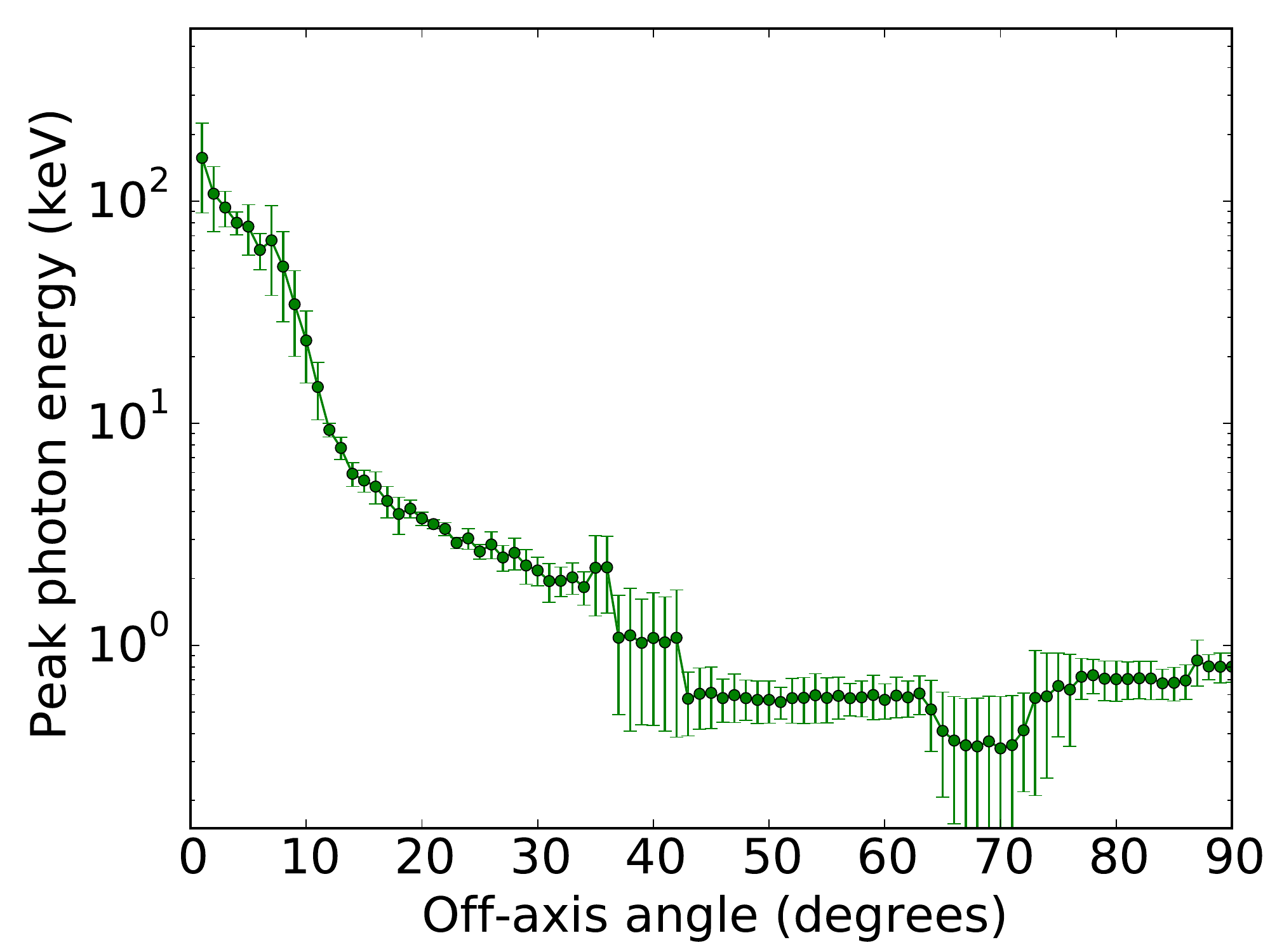}
\caption{{Profile of the peak photon energy of the photospheric
    jet/cocoon emission as a function of the off-axis angle from our
    simulation. The meaning of the symbols and error bars is the same
    as in Figure~\ref{fig:lumi}.}
\label{fig:epk}}
\end{figure}
 
Figure~\ref{fig:epk} shows the profile of the peak energy of the
detected spectrum as a function of the off-axis angle. Also in this
case, we see a three component structure with a very high-energy core
(the on-axis SGRB jet) surrounded by an X-ray dominated cocoon,
eventually transitioning to a predominantly soft X-ray regime at large
angles, with properties independent of the off-axis angle . We notice
that the error bars in the transition regions are particularly large,
indicating that it is where the different components interact that the
assumption of ballistic evolution fails and our results should be
taken with caution. This is particularly obvious at
$\theta\sim40^\circ$, at the cocoon boundary with the shocked ambient
medium, and at $\theta\sim70^\circ$, where the sharp cutoff in
energy is observed.

\begin{figure}
\includegraphics[width=\columnwidth]{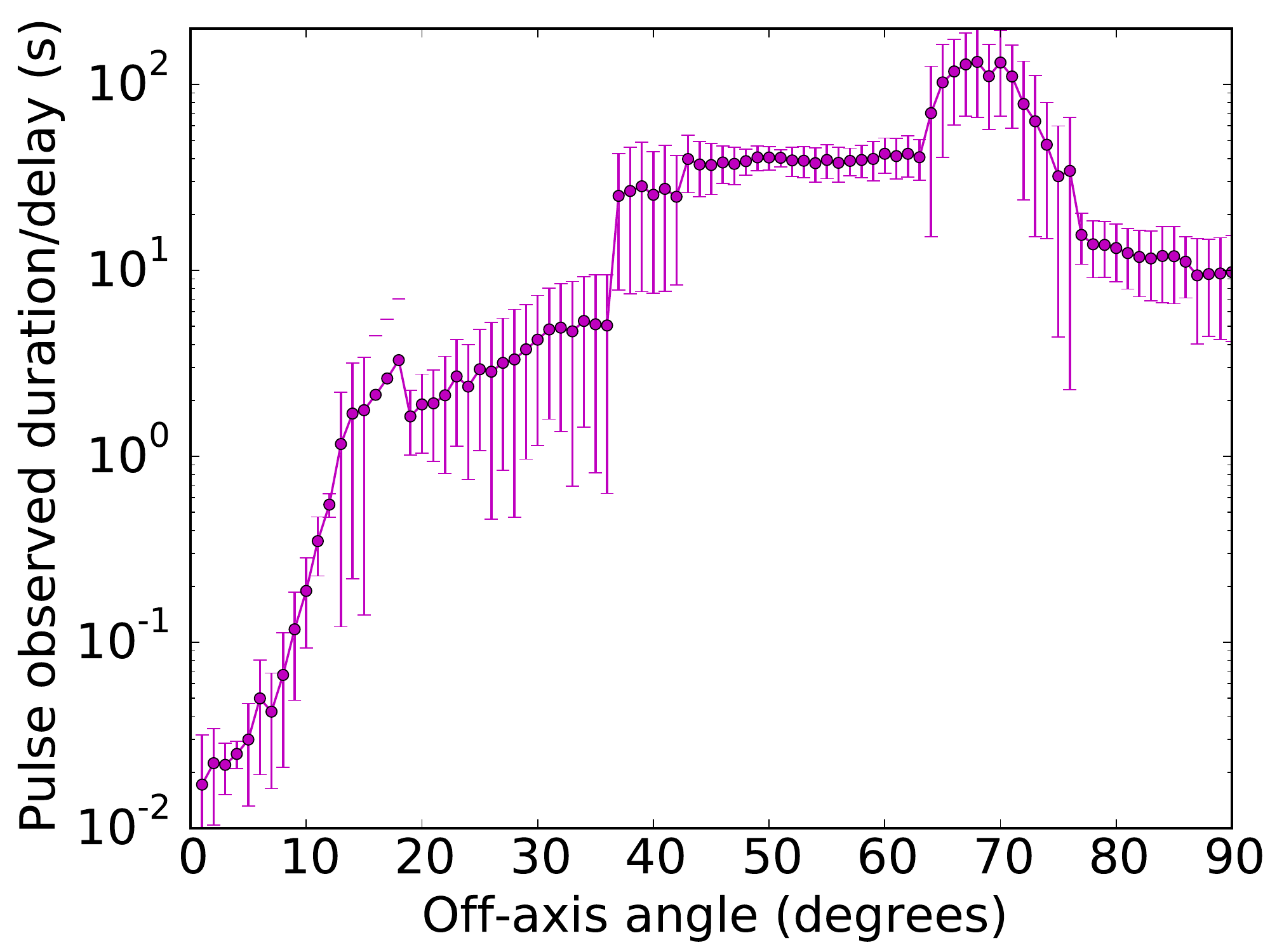}
\caption{{Profile of the pulse duration and time delay of the pulse
    from the jet launching time as a function of the off-axis angle
    from our simulation. The meaning of the symbols and error bars is
    the same as in Figure~\ref{fig:lumi}.}
\label{fig:T90}}
\end{figure}

Figure~\ref{fig:T90} shows the pulse duration as a function of the
off-axis angle. This is computed as the angular time scale
$t_{\rm{ang}}=R/c\Gamma^2$. For the cocoon and shocked ambient medium
material, the angular time scale is the dominant time scale, while for
the jet-dominated case the width of the fireball dominates, and what
is shown may be a lower limit on the duration of the transient. We
notice that the angular time scale coincides with the time delay after
the merger at which the transient is observed, so that a prediction of
this model is that the delay between the engine formation (possibly
indicated by the detection of a GW signal) and the detection of the
transient should be equal to the duration of the transient itself
(Salafia et al. in preparation). We find that this should be of a few
seconds in the cocoon-dominated regime and much longer (minutes) for
the shocked ambient medium.

\begin{figure}
\includegraphics[width=\columnwidth]{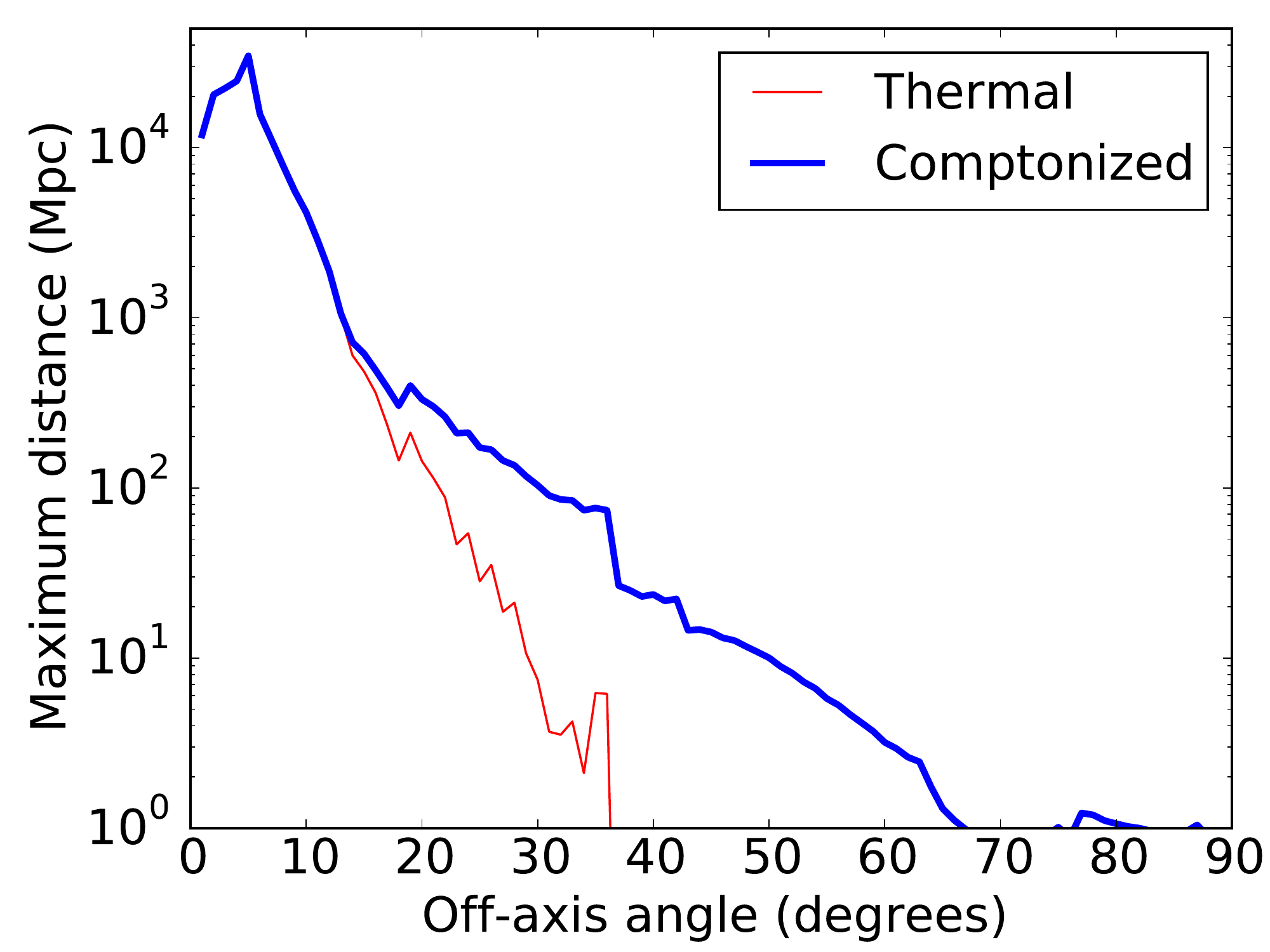}
\caption{{Maximum distance at which the transient can be detected by
    FERMI GBM as a function of the off-axis angle. A pure thermal
    spectrum as well as a Comptonized spectrum are shown. The
    likelihood of detection increases significantly if the spectrum is
    Comptonized by sub-photospheric dissipation, given the low
    temperature of the transient at intermediate and large off-axis
    angles.}
\label{fig:distance}}
\end{figure}

\begin{figure}
\includegraphics[width=\columnwidth]{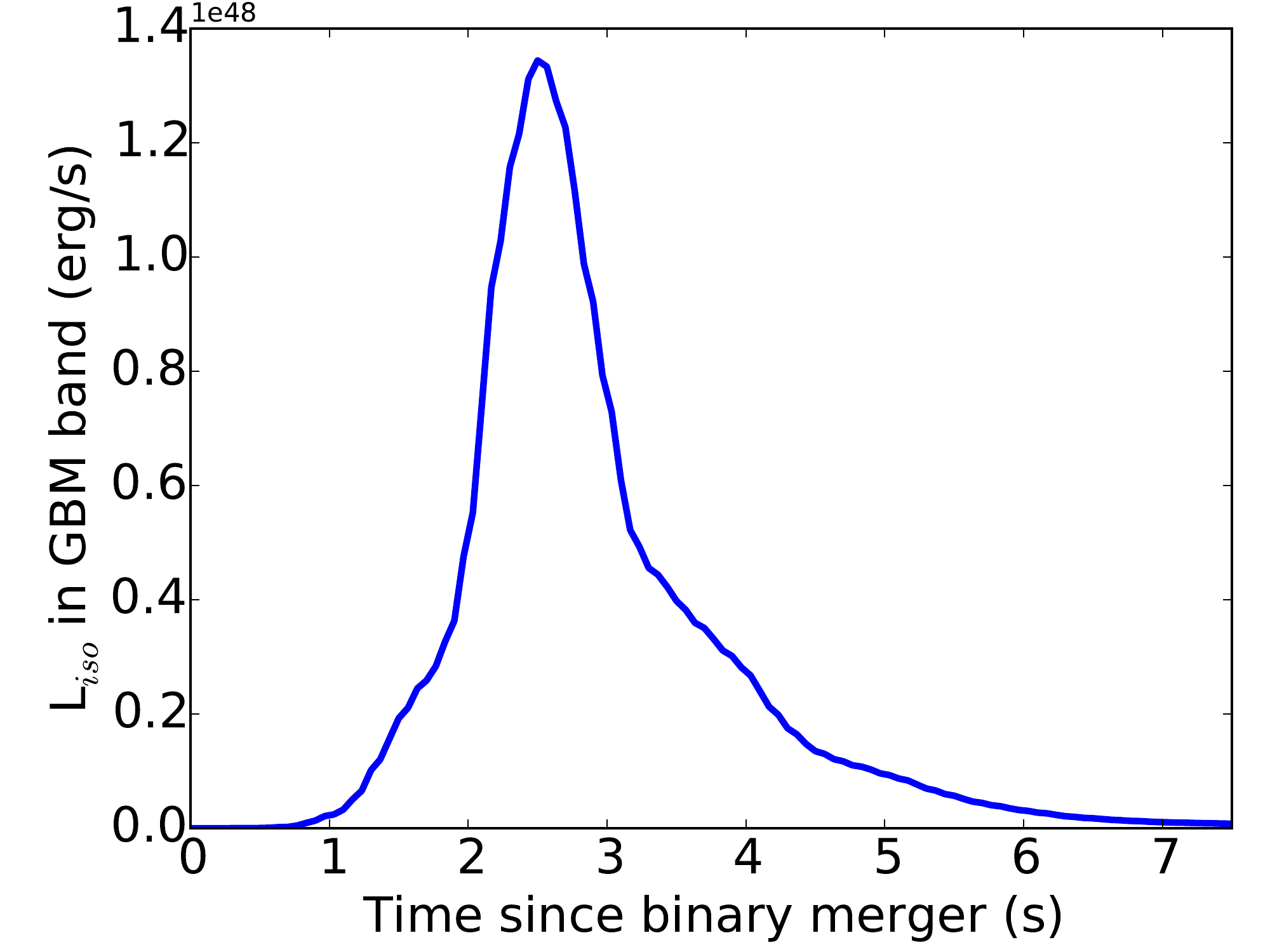}
\caption{{Light curve of the X-ray pulse observed by GBM onboard FERMI
    for a $30^\circ$ off-axis merger. A Comptonized spectrum has been
    assumed. This specific orientation has been chosen since it is the
    one for which the LIGO/Virgo detection probability is maximized
    \cite{Schutz2011}.}
\label{fig:lc30}}
\end{figure}

Finally, we consider out to what distance the prompt EM transient from
a merger observed at a given off-axis angle would be observable with
the GBM onboard FERMI\footnote{Similar distances would be obtained for
  BAT onboard Swift since their sensitivity is comparable for X-ray
  transients like those we discuss.}. Figure~\ref{fig:lumi} shows the
amount of energy that would be seen by the GBM instrument onboard
Fermi. We characterize the BAT sensitivity curve as flat within the
energy range [10-150] keV and zero outside. Since the predicted
photospheric temperatures are at the edge of the sensitivity band of
the instrument, the detected radiation depends sensitively on the
details of the spectrum. We evaluate the detected energy in two
somewhat opposite and extreme cases: a purely thermal, single
temperature spectrum, and a Comptonized spectrum with photon index
$-2.5$\footnote{A spectral index -2.5 is fairly standard for
  high-frequency prompt GRB spectra in the FERMI catalog
  \citep{Gruberetal2014}.}. The latter is obtained by substituting the
part of the thermal spectrum above the peak with the prescribed
power-law. Comptonization of the photospheric spectrum is expected in
case of trans-photospheric
dissipation\citep{Giannios2006,Peeretal2006,Lazzatietal2010,Rydeetal2011,Lundmanetal2013,Chhotrayetal2015}.
To determine the maximum detectable distance, we adopt a transient
detection threshold of 0.7 counts per square cm per second
\citep{Meeganetal2009}. The results are shown in
Figure~\ref{fig:distance}. If the transient is thermal, only fairly
on-axis bursts would be detectable in the $\sim200$~Mpc\footnote{See
  the introduction for a more precise report of the current and
  expected sensitivity ranges of the LIGO/Virgo consortium.} sphere
where LIGO/Virgo are expected to detect NS-NS mergers (red, thin line
in the figure). If, instead, some sub-photospheric dissipation is
present, the cocoon radiation would become detectable at wider
off-axis angles, at least for bursts within $\sim50$~Mpc. In
Figure~\ref{fig:distance}, the thick blue line is computed adding a
high frequency power-law with photon index $-2.5$ on top of the
thermal emission. The extra energy in the power-law photons is, for
all cases, significantly smaller than the energy of the black body
spectrum. It is possible for Comptonization to add significant energy
to the spectrum. We do not consider this case because it would require
either strong shocks (not seen in our simulation) or some form of
magnetic dissipation (which we cannot account for in our AMR-RHD
simulation). An example light curve for the Comptonized pulse observed
by FERMI at $30^\circ$ off-axis is shown in Figure~\ref{fig:lc30}.

\section{Summary and Discussion}

In Lazzati et al. (2017) we discussed the possibility of detecting a
short X-ray transient associated with an off-axis short GRB. The
transient would be due to the expansion of the high-pressure cocoon
that forms around a relativistic jet as it works its way out of a
region of high ambient density. Analogous processes have been
discussed for long GRBs \citep{RamirezRuizetal2002,Nakaretal2017} and
may be responsible for the detection of X-ray flashes
\citep{Gotthelfetal2006,Yamazakietal2002,Fynboetal2004,Lambetal2005,Sakamotoetal2005,Guidorzietal2009}.
In L17, we made some simplifying assumptions, such as the assumption
of isotropy for the cocoon material and the assumption of a Lorentz
factor $\Gamma_{\infty,{\rm{cocoon}}}=10$ for its asymptotic
expansion. Such assumptions (especially isotropy) have been cast in
doubt by \cite{Gottliebetal2017}, who studied the afterglow emission
from such a component (see also L17). In this paper, we have presented
the results of a numerical simulation of a SGRB jet with similar
properties as the fiducial case presented in L17. We find that the
cocoon material produces an X-ray flash detectable by the GBM on board
Fermi under favorable conditions (either a relatively small off-axis
angle $\theta\lesssim30^\circ$ or a distance of tens of Mpc). We also
find that such transients are a few seconds long (compared to a
prediction of a fraction of a second in L17) and peak at a few keV,
making the detection with the GBM somewhat challenging. Qualitatively
analogous results have been presented in
\cite{Kathirgamarajuetal2017}. However, their simulations are
bidimensional and for a magnetized jet, so that our results are not
directly comparable in detail.  It should be noted that neither the
three components classification in Figure~\ref{fig:lumi} nor the
classification as “sheet” discussed in \cite{Kathirgamarajuetal2017}
can capture entirely the complexity of the interaction of a
relativistic jet with the ambient medium through which it
propagates. On the one hand, all components are mixed by reciprocal
interaction and difficult to disentangle. On the other, the presence
of magnetic fields, radiation drag \citep{Chhotrayetal2017}, and
mixing have additional roles in modifying the jet structure and its
surrounding. Further observational work and theoretical studies are
going to be fundamental in the understanding of such a complex and
fascinating phenomenon. The fairly unbiased detection of mergers from
GW detectors with respect of the system inclination will help us
mapping the angular energy distribution of the relativistic and
non-relativistic outflows triggered by the merger (L17).

Contrary to the semi-analytic results in L17, our numerical results
predict X-ray transients from the cocoon that are detectable only
within a range of off-axis angles. Still, the possibility of detecting
at least some off-axis events from their prompt cocoon emission
stands, especially if sub-photospheric dissipation can add a
non-thermal tail to the spectra. It should also be taken into account
that different jet/ambient properties would give rise to different
signals. For example, a less dense environment extended over a larger
volume would produce transients with the same energy but less baryon
contamination and therefore higher temperature (L17).  More luminous
engines active for a shorter time woud instead produce more energetic
cocoons. In extreme cases, the entire jet might be trapped in the
ejecta powering a strong cocoon emission, akin to trapped jets in
massive stars \citep{Lazzatietal2012} and to binary merger engines
that produce outflow with wider collimation \citep{Nagakuraetal2014}.
All such transients would be easier to detect with current
instrumentation.  Finally, we find that the radiative efficiency of
the cocoon is fairly small, of the order of a few per cent, and
therefore the cocoon ejecta energy will be available for producing a
detectable afterglow (see, e.g., L17, Gottlieb et al. 2017).

\acknowledgments This research was in part supported by NASA ATP grant
NNX17AK42G (DL) and NSF awards AST-1616157 (RP) and AST-1333514
(BJM). The software used in this work was developed in part by the DOE
NNSA ASC- and DOE Office of Science ASCR-supported Flash Center for
Computational Science at the University of Chicago. Resources
supporting this work were provided by the NASA High-End Computing
(HEC) Program through the NASA Advanced Supercomputing (NAS) Division
at Ames Research Center.

\end{document}